**Modeling Structural Colors from Disordered One-Component Colloidal Nanoparticle-based Supraballs using Combined Experimental and Simulation Techniques**[†]

Anvay Patil,[1,ϒ,‡] Christian M. Heil,[2,‡] Bram Vanthournout,[3] Saranshu Singla,[1] Ziying Hu,[4] Jan Ilavsky,[5] Nathan C. Gianneschi,[4,6] Matthew D. Shawkey,[3] Sunil K. Sinha,[7] Arthi Jayaraman,[2,8,*] and Ali Dhinojwala[1,*]

[1]*School of Polymer Science and Polymer Engineering, The University of Akron, 170 University Ave, Akron, Ohio 44325, USA.*
[2]*Department of Chemical and Biomolecular Engineering, University of Delaware, 150 Academy St, Newark, Delaware 19716, USA.*
[3]*Evolution and Optics of Nanostructures Group, Department of Biology, Ghent University, Ledeganckstraat 35, Ghent 9000, Belgium.*
[4]*Department of Chemistry, Northwestern University, Evanston, Illinois 60208, USA.*
[5]*X-Ray Science Division, Argonne National Laboratory, Lemont, Illinois 60439, USA.*
[6]*Department of Materials Science and Engineering, Department of Biomedical Engineering, Department of Pharmacology, International Institute of Nanotechnology, Simpson-Querrey Institute, Chemistry of Life Processes Institute, Lurie Cancer Center, Northwestern University, Evanston, Illinois 60208, USA.*
[7]*Department of Physics, University of California San Diego, 9500 Gilman Dr, La Jolla, California 92093, USA.*
[8]*Department of Materials Science and Engineering, University of Delaware, 201 DuPont Hall, Newark, Delaware 19716, USA.*

[ϒ]Current address: *CertainTeed LLC, 20 Moores Road, Malvern, Pennsylvania 19355, USA.*

[‡]A.P. and C.M.H. contributed equally to this work.

*Corresponding authors: ali4@uakron.edu (A.D.), arthij@udel.edu (A.J.)

**Abstract**

Bright, saturated structural colors in birds have inspired synthesis of self-assembled, disordered arrays of assembled nanoparticles with varied particle spacings and refractive indices. However, predicting colors of assembled nanoparticles, and thereby guiding their synthesis, remains challenging due to the effects of multiple scattering and strong absorption. Here, we use a computational approach to first reconstruct the nanoparticles' assembled structures from small-angle scattering measurements and then input the reconstructed structures to a finite-difference time-domain method to predict their color and reflectance. This computational approach is successfully validated by comparing its predictions against experimentally measured reflectance and provides a pathway for reverse engineering colloidal assemblies with desired optical and photothermal properties.

**ToC Figure**

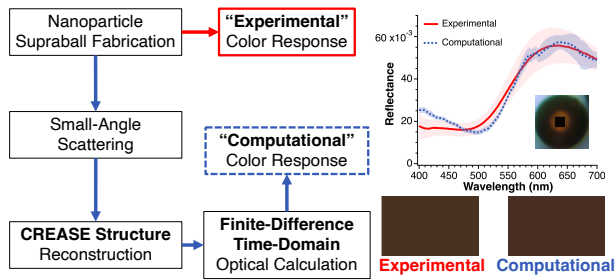

Structural colors in birds and other biological systems[1] have been a source of inspiration for producing synthetic colors for decades.[2,3] Most common synthetic examples have used self-assembly of either polymeric nanoparticles such as polystyrene[4–6] or inorganic nanoparticles such as silica[7–9] to produce disordered colloidal assemblies. Inspired by the chemistry and arrangement of melanosomes (melanin-containing organelles) in bird feathers,[10] researchers have used absorbing nanoparticles such as melanin[3,11] to produce saturated colors. For the most part, the design of these colors has been based on semi-empirical methods by controlling nanoparticle structure (size, dispersity, and packing) and optical properties (complex refractive index).[5,6,12–14] A quantitative approach to model and predict color generation from disordered colloidal assemblies requires knowledge of the internal structure and a robust optical modeling method that handles multiple scattering, large refractive index contrast, and high broadband absorption.

Structural information is typically obtained using electron microscopy or/and small-angle scattering (SAS) techniques.[15,16] Traditional microscopy techniques like electron microscopy only provides structural information as 2D images, whereas more advanced microscopy techniques like tomography can provide a 3D representation. However, these techniques are conducted on a small subset of particles, require long data collection times, necessitate significant data processing, and are limited in the length scale probed compared to SAS.[15,16] On the other hand, SAS techniques, small-angle X-ray or neutron scattering (SAXS/SANS), measure the bulk, ensemble-averaged structural information,[15,16] in which the output of SAXS/SANS experiments is the scattering intensity profile $I$ as a function of the scattering wave vector $q$. Interpretation of the scattering intensity profile is commonly performed using analytical models. However, these models do not provide the 3D structural reconstructions necessary for optical modeling using computational electrodynamic simulations,[15–17] nor do they consider potential structural heterogeneity within the self-assembled films or supraparticles. Additional techniques are thus required for accurate optical analysis.

Optical modeling of densely packed assemblies of absorbing nanoparticles is also challenging. Existing optical modeling tools, including diffusion theory,[18–20] single-scattering approximation based on Mie theory,[5,6,12] and Monte-Carlo-based multiple scattering models,[21,22] treat structures with effective medium approximations, underperform for systems with large refractive index contrasts, and qualitatively capture the trends observed in experiments for barely absorbing materials. For quantitative optical modeling of highly absorbing materials, like melanin, and its dense nanoparticle assemblies, a more direct first-principle technique is needed. The finite-difference time-domain (FDTD) method[17,23] has been shown to predict structural color generation in colloidal nanoparticle assemblies with large refractive index contrasts, high broadband absorption, and dense packing of nanoparticles.[14] However, the use of FDTD requires spatial coordinates of all the nanoparticles within the self-assembled structure.

Here, we present a systematic computational approach to predict structural color generation from disordered colloidal nanoparticle assemblies. This approach combines *a*) the recently developed computational reverse-engineering analysis for scattering experiments (CREASE) tool[24] to reconstruct the 3D structure of assembled particles given an intensity profile from small-angle scattering measurements and *b*) FDTD to calculate the color reflectance spectra from the reconstructed particle assembly structure. We demonstrate that this approach provides simulated reflectance spectra in close agreement with experimental reflectance profiles. This approach presents opportunities to model and predict color generation from complex hierarchical structures that can enable development of structure-color relationships and open exciting avenues to tune structural colors.

As a proof-of-concept, we validate our computational approach using one-component silica (barely absorbing; henceforth referred to as *non-absorbing*) nanoparticle assemblies in a spherical confinement (called supraballs[25,26]) and one-component supraballs with only synthetic melanin nanoparticles (highly absorbing; hereafter referred to as *melanin*). See the Supporting Information for material preparation. The FDTD method requires as input the nanoparticle coordinates (x, y, z) in the nanoparticle assembly under investigation; these nanoparticle coordinates are generated from the CREASE tool which takes as input the small-angle scattering measurement results from the nanoparticle assembly.

We perform SAXS on the primary nanoparticles and supraball systems, as shown in **Figure 1** (see Supporting Information for method details). **Figure 1A** illustrates a conventional transmission geometry of SAXS experiments to obtain a 2D scattering profile that can be averaged to obtain a 1D scattering curve of intensity $I$ as a function of scattering wave vector $q$. Typically, the scattering intensity $I$ of a densely packed assembly, like supraball, can be expressed as proportional to the product of a squared form factor, $F(q)$, term and a structure factor, $S(q)$, term. Prior to measuring scattering profiles from supraball (*i.e.*, assembled nanoparticles at high packing fraction) systems, we determine the primary nanoparticle characteristics called the form factor $F(q)$ (*i.e.*, shape and size) for both silica and melanin nanoparticles (**Figure 1B**). We fit the scattering curves at dilute concentration with a spherical $F(q)$ function for a lognormally distributed polydisperse nanoparticles to obtain the silica and melanin nanoparticle average diameters and dispersities (*silica*: ~244 nm and 3.93%; *melanin*: ~230 nm and 7.29%). These extracted values are supported by a complementary transmission electron microscopy visualization (inset images in **Figure 1B**). Following the form factor measurements, we also collect scattering intensity profiles from one-component silica and melanin supraball (high packing fraction) suspensions to obtain information concerning the nanoparticles' structure (i.e., organization/packing within the supraball geometry) as shown in **Figure 1C**. The internal packing information can be captured by analytically modeling the structure factor, $S(q)$, contribution to the overall scattering intensity. We use a "sticky" hard sphere $S(q)$ model with a Percus-Yevick (PY) closure for an attractive interaction potential[27–29] that results in a qualitative agreement with the experimental scattering profiles. However, the model fit does not output a representative 3D colloidal assembly structure. Instead, we require an approach that produces a structural reconstruction from the scattering datasets to perform optical simulations.

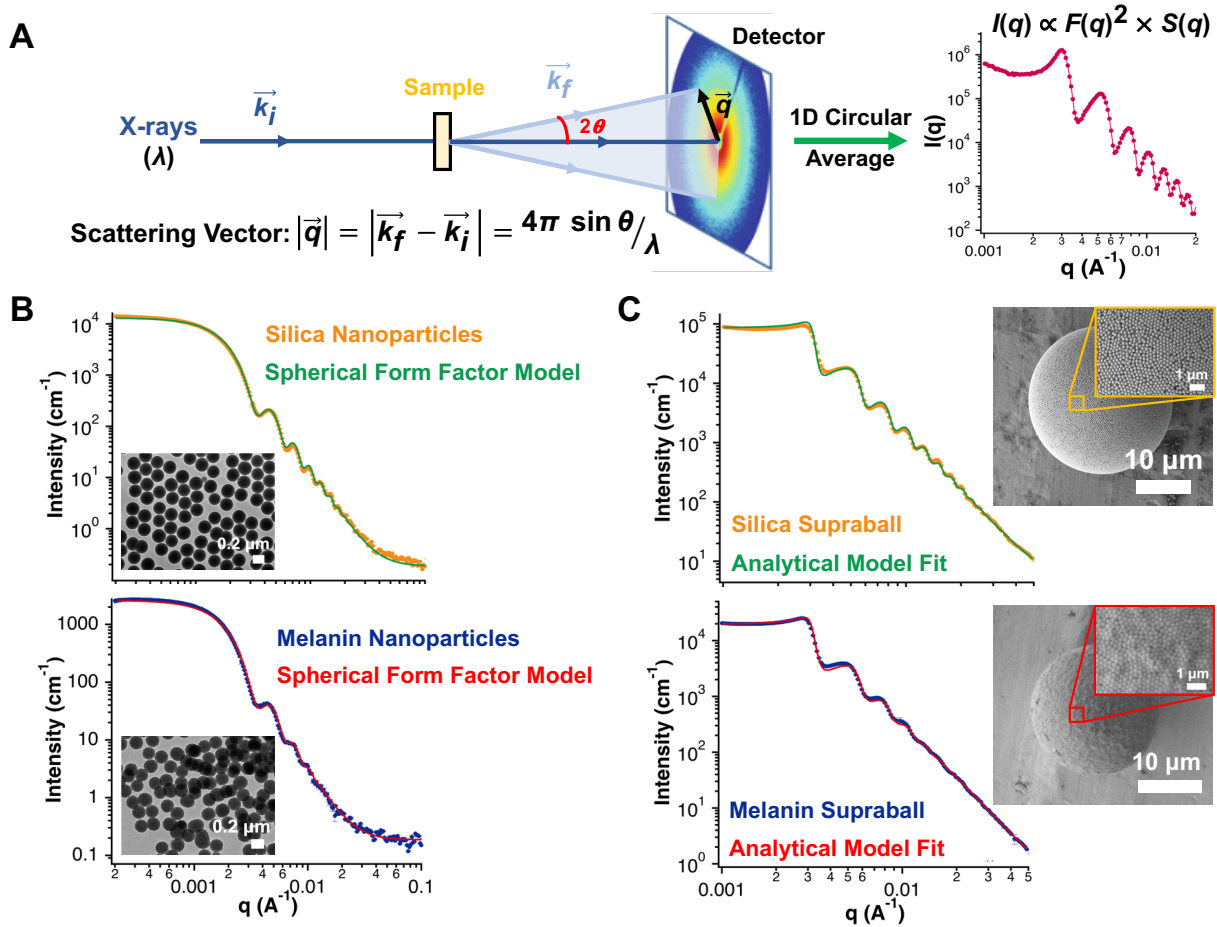

**Figure 1**. *SAXS of melanin and silica nanoparticles and supraballs. (**A**) Schematic of a conventional SAXS geometry in transmission mode. In SAXS experiments, a collimated, monochromatic incident beam of X-rays (represented by a wavevector $k_i$) is scattered at an angle $2\theta$ (represented by a wavevector $k_f$) upon interaction with sample under investigation. These scattered rays are collected by appropriate detection system to record intensity I as a function of scattering wave vector q, given by $|k_f - k_i|$, also represented by $(4\pi \sin\theta)/\lambda$, where $\theta$ is the half-scattering angle and $\lambda$ is the wavelength of X-rays used. (**B**) SAXS plot of I versus q for silica (top; yellow dots) and melanin (bottom; blue dots) nanoparticles in aqueous **__dilute__** suspensions fitted with the corresponding spherical form factor functions for lognormally distributed polydisperse spheres (green curve - silica; red curve - melanin). Inset transmission electron micrographs of silica and melanin nanoparticles used in the experiments. Scale bars are 0.2 μm. (**C**) SAXS plot of I as a function of q for silica (top; yellow dots) and melanin (bottom; blue dots) supraballs fitted with an analytical scattering model representing the structure factor S(q) contribution by "sticky" hard sphere model (green curve - silica; red curve - melanin). Insets are scanning electron micrographs of the supraball (scale bars are 10 μm) and supraball surface (scale bars are 1 μm).*

We apply the recently developed genetic algorithm-based CREASE method[24] to reconstruct the supraball structure in **Figure 2** (see Supporting Information for details). In **Figure 2A**, we show that the CREASE method takes the experimental scattering intensity profile as input and outputs a 3D structure, whose computed scattering profile closely matches the experimental input. The CREASE method utilizes a genetic algorithm, so it stores information representing the nanoparticle structure such as the nanoparticle average diameter and dispersity, nanoparticle concentration, spatial arrangement of nanoparticles, and the number of nanoparticles needed for the 3D structural reconstruction as a set of 'genes'. The genes representing the nanoparticle diameter and dispersity are set using the analysis from **Figure 1B**. We note that the CREASE method can converge those genes even if the mean diameter and dispersity are not known precisely. CREASE then converts those genes into the 3D nanoparticle structure and calculates the computed scattering intensity $I_{comp}(q)$ with the Debye scattering equation[30,31] or using machine learning.[24] CREASE calculates the fitness for each individual which quantifies the quality of the match between $I_{comp}(q)$ and $I_{expt}(q)$. After each generation, individuals' sets of genes are altered by performing mutations and combinations that randomly change the gene or average two good gene values together, respectively. After the fitness converges, CREASE outputs the best 3D nanoparticle structure (best set of genes) and its corresponding $I_{comp}(q)$ (which closely matches the $I_{expt}(q)$) for use in the FDTD calculations.

For the gene related to number of nanoparticles needed for structural reconstruction, the number of nanoparticles in the reconstruction to generate a 3D reconstructed structure with a similar dimension as the experimental supraballs (diameter ~10 μm) would require ~65,000 nanoparticles. Using this large number of nanoparticles in every step of the genetic algorithm in the CREASE method would be computationally and time intensive to determine the remaining gene values that correspond to a structure with a computed scattering intensity, $I_{comp}(q)$, that most closely matches the target *i.e.*, experimental scattering input, $I_{expt}(q)$. Instead, we leverage the gene-based nature of CREASE to optimize the gene values for a reconstruction of diameter ~6 μm using a smaller number of nanoparticles (~20,000 nanoparticles) during optimization.

After we have optimized the other gene values related to the spatial arrangement of nanoparticles, we reconstruct a 3D structure with a large number of nanoparticles to achieve similar dimension to the experimental supraballs. This two-step approach reduces the computational time required while obtaining output structures with strong agreement between the $I_{expt}(q)$ and $I_{comp}(q)$ for both silica (*top*) and melanin (*bottom*) in **Figure 2B**. We note that it is difficult to distinguish between the $I_{expt}(q)$ and $I_{comp}(q)$ for the silica and melanin supraballs due to how closely the two scattering profiles match, and that close scattering match means the CREASE output is structurally similar to the experimental supraball. Additionally, we also confirm that the gene representing the nanoparticle concentration converges to the correct value, indicating the structure is entirely one chemistry of nanoparticles as expected for the one-component supraballs. In **Figure 2C**, we compare the $S(q)$ calculated from the CREASE reconstructed structures using the Debye equation[30,31] to that obtained from the analytical model fit to the experimental supraball scattering intensity curves (**Figure 1C**). The $S(q)$ calculation using the Debye equation directly accounts for the nanoparticle size distribution while the "sticky" hard sphere $S(q)$ model assumes monodisperse nanoparticles. This causes the CREASE $S(q)$ to be closer to the experimental $S(q)$. The monodisperse assumption of the analytical model leads to narrower peak widths and larger peak heights compared to the CREASE $S(q)$. Despite the difference in assumptions, the $S(q)$s possess similar shape and features especially in the low $q$ regime, corresponding to large distances in real space, for both the silica and melanin chemistries.

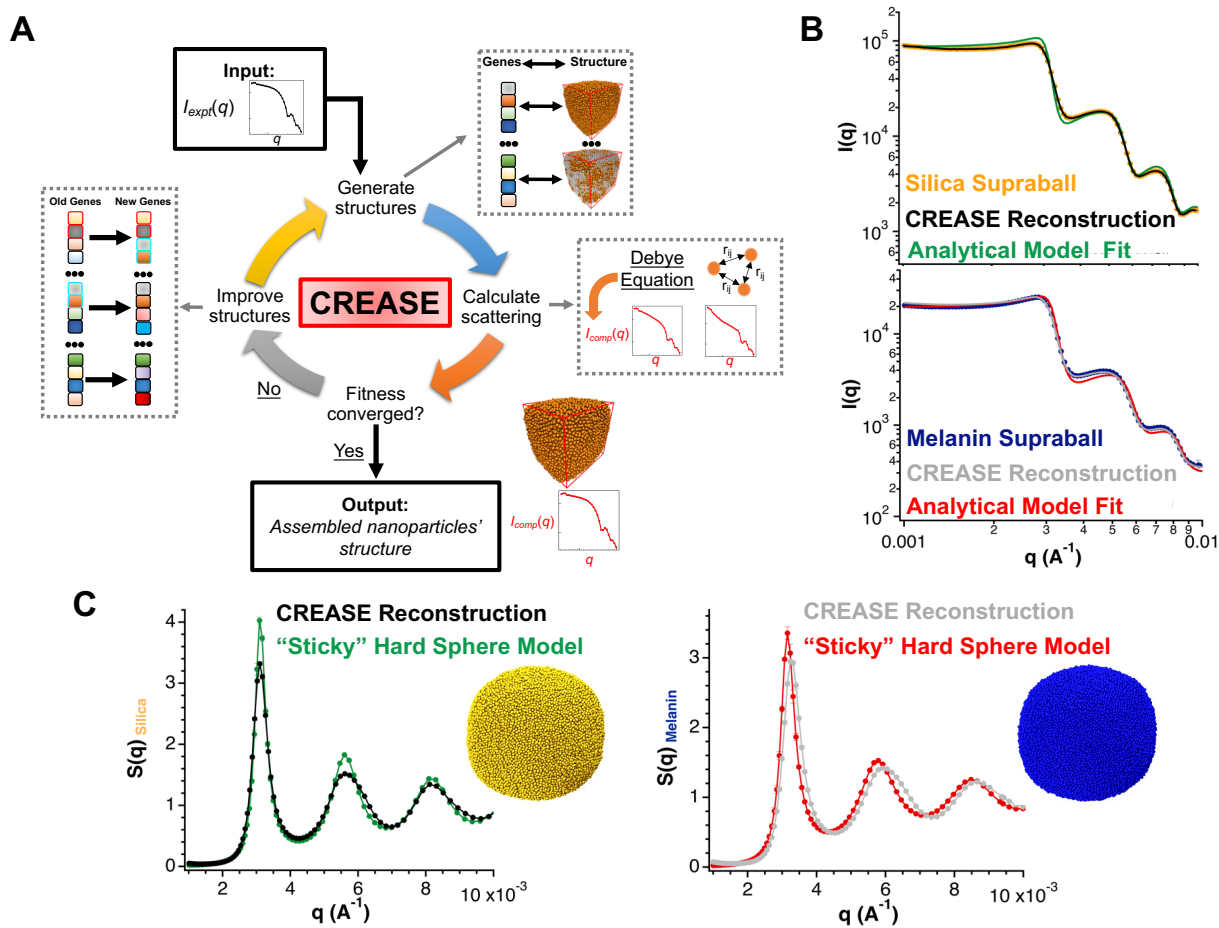

**Figure 2**. *Applying the CREASE method to reconstruct the nanoparticle assembly structure from SAXS profiles. (A) Schematic describing the CREASE method operation. CREASE requires a scattering profile as an input and generates a representative 3D structure as an output. (B) SAXS plot of I as a function of q for silica (top; yellow dots) and melanin (bottom; blue dots) supraballs overlaid with the CREASE output structures' scattering profile for silica (top; black line) and melanin (bottom; gray line) and the analytical scattering model for silica (top; green curve) and melanin (bottom; red curve). (C) A comparative plot of structure factor, S(q), between that calculated from the CREASE reconstruction for silica (left; black) and melanin (right; gray), and that from the 'sticky' hard sphere S(q) model for silica (left; green) and melanin (right; red). Insets are VMD visualizations of the reconstructed 3D colloidal nanoparticle assembly with yellow spheres representing silica chemistry (left) and blue spheres representing melanin (right).*

In **Figure 3**, we show results from optical modeling using the FDTD method on the CREASE's output 3D structures and coordinates to calculate the computed reflectance spectra (see Supporting Information for method details). The FDTD method provides a general solution to all light scattering problems. It determines the reflectance, transmittance, and absorbance of the supraballs by discretizing the supraball geometry onto a spatiotemporal grid (mesh) to iteratively solve Maxwell's curl equations numerically at each grid point and at each time step. Thus, FDTD can directly capture relevant supraball structural characteristics (like nanoparticle size, size dispersity, and packing), record the time evolution of electric and magnetic vector field components as light traverses through the structure, and provide the requisite optical response of the system under investigation. To perform FDTD calculations on the supraball system, one requires the relevant material properties (complex refractive index) and the x, y, and z positions of the nanoparticles from the CREASE reconstruction.

While **Figure 2** suggests that the CREASE output structures possess similar structural attributes as the experimental supraballs based on the scattering intensity profiles, FDTD enables a separate, additional structural validation on the CREASE output. For the non-absorbing silica supraballs in **Figure 3A**, the computed reflectance spectrum closely matches the experimental reflectance. The visual perception of the computed reflectance, shown by the RGB color panel, agrees with the experimental RGB color. We also provide an optical micrograph of a silica supraball as a comparison to the RGB color panels. The chromaticity coordinates of both experimental and computed colors are marked on the CIE 1931 color chart to indicate their relative closeness. Color difference ($\Delta E$) analysis reveals that the computed and experimental colors are similar, only ~1.1 times the average just noticeable difference (JND) value,[32] indicating that barely more than 50% of observers can distinguish between them. The color and reflectance spectra matches are even more impressive considering that the CREASE method is an open-source, generic, reconstruction tool not specially designed or modified for application to these supraballs. The quantitative color match between experiments and computations is also high for the absorbing melanin supraballs in **Figure 3B**. The computed and experimental reflectance spectra are largely indistinguishable within error. The slight reflectance mismatch particularly at low wavelength may be caused by the ordered supraball surface (see inset in **Figure 1C**) compared to the less ordered, bulk CREASE reconstruction.[14] **Figure 3B** provides both quantitative and visual comparisons for the colors obtained from the experimental measurement and optical modeling of the CREASE output structures. The two colors are similar, with their $\Delta E$ only ~2.1 times the average JND value. By further confirming the CREASE output structural match using optical modeling, we have higher confidence that the reconstructed structures mimic the bulk experimental structure including the relevant color-producing structural components. This combined computational approach opens an avenue to study structure-color relationships that can be leveraged to modulate light reflectance in specific wavelength bands.

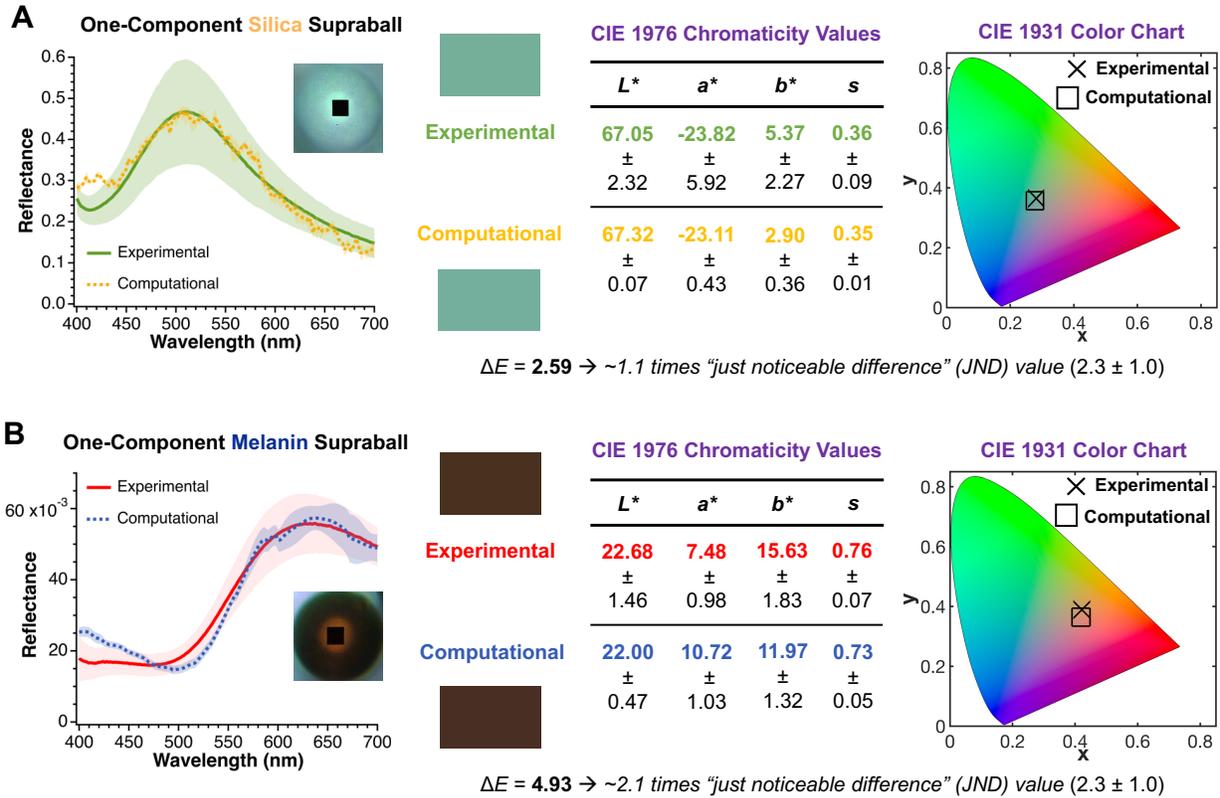

**Figure 3.** *Optical modeling comparison between the computed reflectance spectra obtained from FDTD calculations on the CREASE output structures and the experimental reflectance profiles. (**A**) Reflectance spectra (green solid curve: experimental; yellow dashed curve: computed), RGB color panel, CIE 1976 chromaticity values, and CIE 1931 chromaticity co-ordinates' comparisons for the non-absorbing silica supraball system. The quantitative difference between FDTD and experimental colors is given by a color difference (ΔE) value that is ~1.1 times the average just noticeable difference (JND) value. (**B**) Reflectance spectra (red solid curve: experimental; blue dashed curve: computed), RGB color panel, CIE 1976 chromaticity values, and CIE 1931 chromaticity co-ordinates' comparisons for the highly absorbing melanin supraball system. The quantitative difference between FDTD and experimental colors is given by a ΔE value that is ~2.1 times the average JND value. The black box in the inset of the optical micrographs of the corresponding supraballs (**A** and **B**) represents the size of the area (3 μm × 3 μm) probed during optical measurements using microspectrophotometer.*

We also investigate the CREASE output structure size (by changing the number of nanoparticles used in the method) to determine if adjusting the reconstructed structure size influences the reflectance spectra. For both the non-absorbing silica and absorbing melanin supraballs, the reflectance spectra shape is consistent when we consider both smaller (~6 μm; ~20,000 particles) and larger (~20 μm; ~250,000 particles) reconstructions (**Figure S1**). Interestingly, for the silica system, the CREASE reconstruction size influences reflectance intensity with the larger (smaller) structures, having more (fewer) silica nanoparticles, allowing for additional (fewer) multiple scattering events that ultimately increase (decrease) the reflection. On the other hand, the melanin-based CREASE structures exhibit quantitatively similar reflectance values within the simulation error regardless of CREASE structure size. Because melanin is an absorbing nanoparticle, increasing the CREASE reconstruction size beyond a certain critical value does not further reduce reflectance, as a sufficient number of melanin nanoparticles to absorb as much incident light as possible during multiple scattering events is already present. For both chemistries, we note that the larger CREASE reconstructions achieve closer SAXS intensity profile matches to the experimental systems because the low $q$ (0.0001 to 0.001 $A^{-1}$) scattering intensity is dominated by the large form factor description of experimental supraballs.

Before concluding this work, we provide pertinent information on utilizing the CREASE-FDTD approach and potential avenues for future work. For a researcher to utilize this CREASE-FDTD approach on a similarly sized system of interest (~65,000 nanoparticles), they would require ~1 hour on a single core for the machine learning CREASE method to converge the genes[24] and generate the 3D nanoparticle structure for those converged genes and ~8-10 hours on 14 cores for the FDTD calculation on the reconstructed structure. Thus, in less than half a day, a researcher could determine their material's structure and validate that the reconstructed structure possesses the desired optical properties, enabling rapid development of structure-color relationships. Because this approach is computational, one can perform this process on multiple different systems at once, if one had a sufficient number of computational resources, enabling high-throughput analysis!

Furthermore, this approach possesses the potential for use in the development of related materials with targeted optical properties. Once CREASE converges to the genes for a particular material, a researcher could easily adjust one or more genes to create a new structure and perform FDTD on the new structure to determine the new material's optical properties. For example, researchers could adjust the gene corresponding to the nanoparticle diameter, nanoparticle dispersity, the nanoparticle composition/chemistry (multi-component materials), *etc*. and predict the resulting material color. Once the color fine-tuning has been performed, experimental synthesis of the system would be used to corroborate the most interesting computational results, saving significant resources compared to experimentally examining all new materials. This approach promises to help researchers tailor the material to achieve a specific coloration, and with our previous success in using machine learning to link genes to scattering profiles,[24] we believe a machine learning approach linking genes to color would further reduce resource costs and structure optimization time. Finally, while this work has focused on one-component supraballs, this CREASE-FDTD approach is not limited to single chemistry nanoparticle assemblies. Future work involving multi-chemistry supraballs would further demonstrate the potential applicability of this combined computational approach.

In summary, this work demonstrates a proof-of-concept computational method to predict structural colors from colloidal nanoparticle assemblies using a combination of experimental structural reconstruction via the CREASE method and optical modeling via the FDTD toolbox. For both melanin and silica systems, the CREASE and FDTD approach produces reconstructed nanoparticle assemblies with similar color reflectance properties as that seen with the experimental systems. This two-part computational approach will enable researchers to design more complex heterogeneous or multi-component self-assembly of nanoparticles of different shapes, sizes, and chemistries to tune visible colors or spectral response in other regions of electromagnetic wavelengths.

**Supporting Information:**

- Detailed experimental and computational methods and materials. Information on color analysis using CIE standards. Additional results on the effect of CREASE reconstruction size on computed reflectance, demonstration of silicon substrate effect on reflectance, extinction coefficient for silica nanoparticles, and CIE standards used in calculation of tristimulus values.

**Acknowledgements:**

A.P., C.M.H., S.S., Z.H., N.C.G., A.J., and A.D. acknowledge financial support from the Air Force Office of Scientific Research (AFOSR) under Multidisciplinary University Research Initiative (MURI) grant (FA 9550-18-1-0142). S.K.S. acknowledges support from the Office of Basic Energy Sciences, U.S. Department of Energy (DOE) under DOE grant No. DE-SC0018086. B.V. and M.D.S. acknowledges support from AFOSR grant (FA9550-18-1-0477), FWO grant (G007117N), and HFSP grant (RGP 0047). This research used resources of the Advanced Photon Source (APS), a U.S. DOE Office of Science User Facility, operated for the DOE Office of Science by Argonne National Laboratory (ANL) under Contract No. DE-AC02-06CH11357. A.P., S.S., J.I., S.K.S., and A.D. thank the efforts of Ivan Kuzmenko during data collection at the APS 9-ID beamline. Finally, we extend our gratitude to the entire MURI Melanin team for helpful discussions and insights throughout the course of this project.

**Author Contributions:**

‡A.P. and C.M.H. contributed equally to this work. A.P., C.M.H., A.J., and A.D. contributed to project conceptualization and design of experiments. A.P. performed scattering experiments, optical modeling (FDTD calculations), and data analysis. C.M.H. conducted CREASE development and structure reconstructions. Z.H. and N.C.G. performed synthesis of melanin and silica nanoparticles. A.P. and S.S. fabricated one-component melanin and silica supraballs. B.V. and M.D.S. assisted in the reflectance measurements of one-component melanin and silica supraballs. A.P., S.S., and J.I. performed X-ray scattering experiments at the APS 9-ID beamline, processed raw data, and completed data fitting. S.K.S. provided guidance during X-ray scattering experiments, augmented conceptual understanding, and assisted in data processing and fitting. A.P. and C.M.H. wrote the initial manuscript; A.P., C.M.H., A.J., and A.D. reviewed and edited the manuscript. All authors have given approval to the final version of this manuscript.